\documentclass[10pt,letterpaper,twocolumn]{article}
\usepackage{graphicx}

\begin{document}

%\preprint{HEP/123-qed}

\title{Exploring closed-loop feedback control \\ using experiments in optics}

\author{
K. A. Jensen, R. J. Larson, and S. D.
Bergeson\thanks{scott.bergeson@byu.edu} \\ Department of Physics
and Astronomy, \\ Brigham Young University, Provo, UT 84602
\and
E. F. McCormack\thanks{emccorma@brynmawr.edu} \\ Department of
Physics,  \\ Bryn Mawr College, 101 N. Merion Ave., Bryn Mawr, PA
19010-2899}
\date{\today}

\maketitle

%\clearpage

\begin{abstract}
We present two experiments in closed-loop feedback control.  In
the first experiment, students control the pointing angle of a
laser to ``lock'' the laser onto a ``target.''  In the second,
students stabilize the pathlength difference in two arms of a
Michelson interferometer.  These experiments are appropriate for
electronics and optics laboratory classes for junior and senior
level students.
\end{abstract}

\maketitle

\section{Introduction}

Feedback control systems are all around us.  Cruise control,
climate control, temperature regulation in engines, pressure
control in industrial steam systems, power supply current and
voltage regulators, frequency controllers and many other systems
use feedback control. Our own bodies use this kind of control to
stand and walk, to watch a passing object, and to regulate body
temperature.

Your home or office climate-control system, for example, uses
feedback to maintain an inside temperature of
$23~^{\mbox{\footnotesize{o}}}$C (your set point). On a cold
morning, the heater turns on.  When the temperature rises above
the set point, the heater turns off and the room gradually cools.
Eventually the temperature falls below the set point, and the
heater turns on again.

This particular kind of automatic feedback control is relatively
crude.  It falls into the class of ``proportional controllers,''
where the feedback signal is proportional to the error signal.  In
this example, the temperature is not maintained exactly at the set
point, but oscillates within a small range around it.  As long as
the oscillation is not too great, the temperature feels constant,
and the control system is adequate.

However, many applications require tighter regulation, and the
feedback systems become more complicated.  Better regulation
requires detailed information about the response time of the
system to a given input, the size and frequency content of the
perturbing signals, and the necessary regulation tolerances for
the system.

Several textbooks describe generalized approaches for implementing
feedback control (see, for example, \cite{Dorf}).  These
treatments typically use mathematical models, differential
equations, Laplace transforms, complex algebra, and computer
simulations to design the optimum feedback system with the proper
gain, roll-off frequency, and phase margin. The strength of these
treatments is that they can be applied to all kinds of systems,
from tiny voltage control circuits to large manufacturing
operations and complex delivery systems.

For many students, these powerful and generalized treatments of
feedback control are difficult to penetrate.  Somehow the simple
physical picture of stabilizing your system output is lost in
mathematical models and in the jargon of the discipline.

This paper is an attempt to provide a few simple hands-on
illustrations of feedback control in a laboratory setting.  In it
we walk the reader through two brief and relatively inexpensive
implementations of feedback control.  The presentation is fairly
conceptual.  It does not include a mathematical model of the
systems.  Also missing is a detailed discussion of system
stability and the associated methods of modeling, measuring, and
implementing high stability control systems.  However, we do point
out departure points in the discussion where interested readers
can find more information in the literature.  In our opinion, this
introductory presentation gives students (and instructors) an
intuitive introduction to the basic concepts of feedback control.

\section{Laser Tracking Device}

%The basic idea of feedback control is simple.  A system of some
%kind needs to produce a particular output.  Typically, the system
%is turned on and the output is compared to the desired output, or
%``set point.''  The difference between the desired and actual
%outputs is called an ``error signal.''  Some kind of feedback
%apparatus uses this error signal to change the system operating
%parameters and bring the system output closer to the ideal.

Our first feedback control example is a laser tracker---a generic
control system.  With variations in the pointing device and the
detector, the same basic idea is used for missile tracking systems
and telescope pointing stabilization.  In this lab, students build
an automatic pointing system that keeps a laser pointing at the
center of a ``target.''

\begin{figure}
\includegraphics[angle=270,width=3.4in]{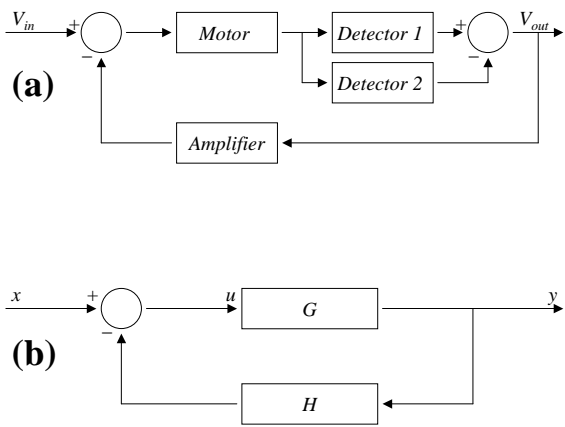} %figure 1
\caption{Block diagram of the laser tracker experiment.  (a)
Detailed diagram.  An input voltage ($V_{in}$) drives the motor.
The difference of the two detector signals is the output voltage
($V_{out}$), or in our case, the error signal. This error signal
is amplified and added to the input voltage to change the pointing
of the laser.  (b) Traditional block diagram. The system input is
$x$, and the output is $y$. The function $G$ represents the
transfer from voltage in to voltage out, sometimes called the
forward gain of the system.  The function $H$ is the feedback
gain.} \label{fig:laserblock}
\end{figure}

The block diagram in Figure \ref{fig:laserblock} outlines the
major portions of the experiment.  We mount the small key-chain
laser-pointer directly on the shaft of a DC motor.  The detector
is a two-segment photodetector that our students build, as
described below.

In Figure \ref{fig:laserblock}a, the voltage $V_{in}$ at the motor
changes the laser's pointing angle. If the laser points at the
center of the target, the signal out from the two detectors is
identical. The difference of the two detector signals, $V_{out}$,
is zero. If the laser points  a little to the right, $V_{out}$ is
negative (say); and if the laser points a little to the left,
$V_{out}$ is positive.  The voltage $V_{out}$ is amplified and
added into the input signal to move the laser back toward the
center of the target.  This is the essence of closed-loop feedback
control: the system output is sampled and compared to the desired
output; and the system input is changed in such a way as to bring
the system output closer to the desired output.

A more traditional block diagram, typical of those used in
feedback-control analysis, is shown in Figure
\ref{fig:laserblock}b.  In the Figure, $G$ represents the
``forward gain'' or open loop transfer function that converts the
input voltage into the output voltage.  For small angular
variations about the center of the target, the voltage output $y$
is linearly proportional to the angular error, and passes through
zero at zero angular error.  (In other words, the system is linear
and homogeneous in a mathematical sense.)  The output $y$ is
amplified by the block $H$ and subtracted from the input $x$.  The
variables $x$, $y$, and $u$ are related by the equations

\begin{eqnarray}
y & = & Gu \\
u & = & x - Hy.
\label{eqn:transfer}
\end{eqnarray}

\noindent Solving the above equations, the input and output for
the system are related by the equation

\begin{equation}
y = \left( {{G}\over{1 + GH}} \right) x.
\end{equation}

\noindent The term inside the parentheses is called the
closed-loop transfer function for the system.

In the typical case where $GH \gg 1$, the closed loop transfer
function is $y \approx Gx/GH = x/H$.  So for large values of the
feedback gain $H$, the output ($V_{out}$ in the laser tracker
experiment) goes to zero, meaning that the laser points exactly at
the center of the target.

This analysis demonstrates that tight regulation of the output
requires very high feedback gain.  Of course there are practical
problems with high feedback gain, depending on the particular
system under study.  For example, in mechanical systems when the
feedback gain is high, the system responds quickly to
perturbations.  But if the gain is too high, the electronics can
overdrive the motors, and damage gears or shafts.

There is another problem with very high feedback gain.  In the
above analysis we implicitly assume that the system can respond
instantaneously to the correction signal.  If there is a time
delay between when the error signal is measured and when it is
corrected, either because of mechanical or electrical limitations,
then too large a feedback gain can send the system into
oscillation.

In a more sophisticated analysis of the system, the forward gain
$G$ and the feedback gain $H$ are both functions of frequency. The
frequency-dependent gains and their associated phase shifts can be
measured by driving the system with sinusoidally varying inputs of
different frequencies, then comparing the input and output.  With
gain and phase shift information, students can model their
transfer functions as either high-pass, low-pass, or band-pass
filters. Equation \ref{eqn:transfer} becomes a differential
equation.  The poles and zeros of the equation, which can be found
using Laplace transforms and associated mathematical analysis,
indicate natural resonances in the system. Obviously, the poles
can cause problems unless the feedback gain is sufficiently low at
those frequencies. It is possible to suppress the influence of the
poles in $G$ by adding zeros in $H$ at those frequencies. For more
information, the reader is referred to the literature \cite{Dorf}.

\subsection{Some Machining}

There are, of course, several possible implementations of this
experiment. In our class, an introduction to the machine shop is
an important segment, and this project is a nice way to do it.  We
build our own target, which is a two-segment photo-detector.  Our
target could be replaced by a split photodiode, such as sold by
Hamamatsu, UDT, or any number of other manufacturers.

\begin{figure}
\includegraphics[angle=270,width=3.4in]{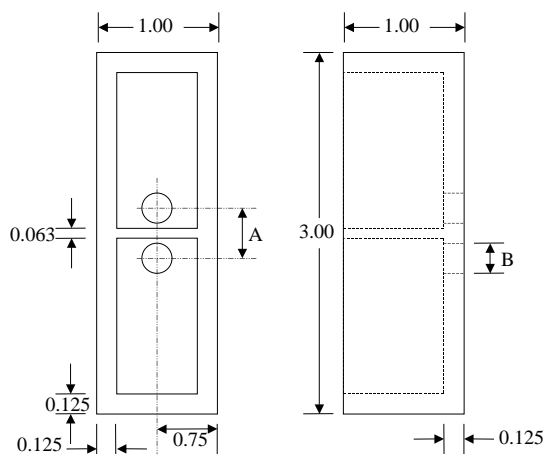}%figure 2
\caption{Drawing of our detector.  Students machine this out of a
single piece of aluminum.  The phototransistors are mounted in the
small holes in the back of the detector.  The dimension {\bf A}
can be as small as conveniently possible.  The dimension {\bf B}
should be sized to your detector.  The dimensions are given in
inches. A frosted glass microscope slide is epoxied to the front
to disperse the light in the detector.} \label{fig:detector}
\end{figure}

Figure \ref{fig:detector} is a drawing of our detector---an
aluminum box three inches long, one inch wide, and one inch deep.
The box is hollowed out so it looks roughly like a square-ish
canoe.  A thin aluminum plate running the short way across the
middle of the box divides the detector into nearly square right
and left chambers of equal size.  The box is hollowed out of a
single piece of aluminum.  We epoxy two phototransistors through
holes on the bottom of the box, one in each chamber near the
dividing plate. Actually, the dividing plate is probably
superfluous, but it is conceptually nice to have the two detectors
physically separated.

We epoxy a sand-blasted microscope slide to the top of the
detector, {\sl the rough side facing out}.  The slide diffuses the
laser light in the detection chambers, making it possible to
detect the laser using the phototransistor when the laser is not
pointing directly at the transistor.  It also serves the important
safety role of eliminating specular reflections from the detector.

The laser pointer is mounted on the shaft of 12V DC motor, using a
small aluminum block machined by the students.  Our motor turns
too quickly even at the lowest operating voltage, so we step the
rotation speed down 10x with a pair of gears.  The motor is not
ideal for tight control because of a hysteresis problem.  The
rotation speed is proportional to the drive voltage until the
voltage drops below 2V, when the motor abruptly stops.  The motor
does not turn again until the drive voltage drops below -2V, when
it starts rotating in the other direction. In spite of this,
feedback control still works to keep the laser pointing at the
center of the detector with a small offset error.

\subsection{Electronics}

\begin{figure}
\includegraphics[angle=270,width=3.35in]{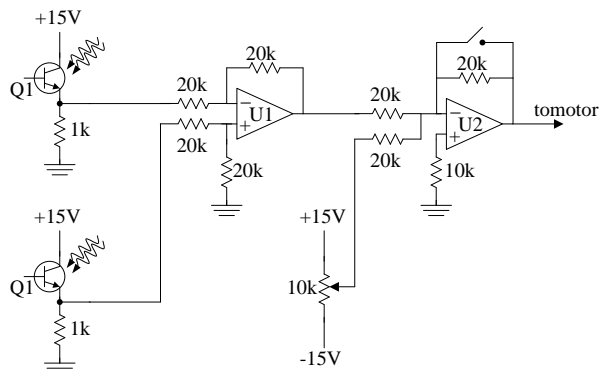}%figure 3
\caption{Schematic diagram for laser tracker control circuit. Q1 =
phototransistor (MR370), U1 = operational amplifier (AD823AN), U2
= high-current op-amp (LM6171).} \label{fig:laserschematic}
\end{figure}

Figure \ref{fig:laserschematic} is a schematic diagram for the
electronics used in this experiment.  As shown, the
phototransistor collector is wired to +15V.  The base is unwired,
and the emitter is connected to ground through a 1 k$\Omega$
resistor.  The current emitted by the phototransistor is linearly
proportional to the incident light intensity.  The 1 k$\Omega$
resistor converts that to a voltage.  The value of this resistor
is chosen to keep the current emitted by the phototransistor below
the maximum current rating for the device.

The output of U1 is the difference in the voltages out from the
two phototransistors.  In our setup, U1 is a standard operational
amplifier that can source only 10 to 20 mA of current.  Because
our motor draws 120 mA, we use U2 (the LM6171) as a high-current
buffer.  Also shown in the schematic diagram is a variable voltage
going into U2.  This is intended to provide some control of the
laser pointing angle when the system is not locked \cite{lock}.
There is a ``capture range'' for the feedback circuit: the laser
needs to point near the middle of the target in order for the
control to work.  This extra control of the laser-pointing
direction allows the students to move the laser into the capture
range.

The output of U1 is called the ``error signal,''  the signal that
indicates how far from the center of the detector the laser
points, and which direction (left or right).  In the jargon of
feedback control, the amplifier U2 is called the ``controller.''
In the present case, our controller has only proportional gain.
Other kinds of feedback controllers use integral gain or
differential gain or combinations of the three.

\begin{figure}
\includegraphics[angle=270,width=3.4in]{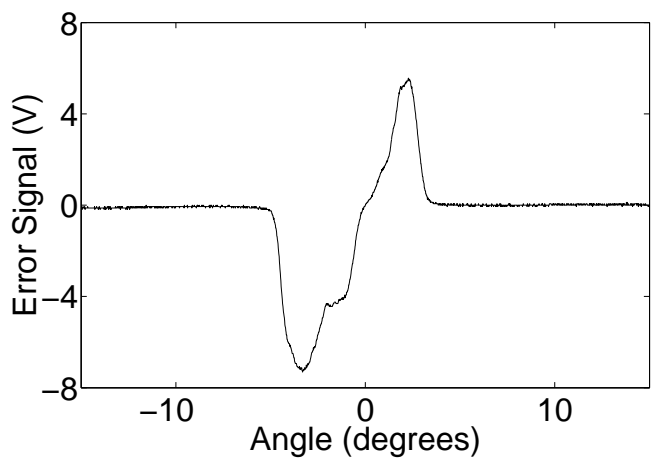} %figure 4
\caption{Error signal from our laser tracker.  Even though it is
not quite symmetric, due to the different sensitivities of our two
detectors, it is roughly dispersion-shaped.  Near the center, the
error signal is linear, and passes through zero volts at zero
pointing error.} \label{fig:errorsignal}
\end{figure}

An error signal generated in our setup is shown in Figure
\ref{fig:errorsignal}.  This is the output from U2 (the
controller) as the motor sweeps the laser across the face of our
detector.  Notice that when the laser points directly to the
center of the detector, the error signal is zero, making the
system homogeneous in the mathematical sense. Also notice that the
slope of the error signal is approximately linear, and has a value
near 2 Volts/degree.  The output of U2 goes directly to the DC
motor through a 120 Ohm, 2 Watt resistor.  This resistor is chosen
to prevent overdriving U2, which at 15 V can source only 0.13 A.

\subsection{Putting it Together}

Now you are ready to put the system together.  To make the laser
lock onto the target, you first need to get the laser pointing in
approximately the right place with the feedback system disabled.
Typically, we do this by first shorting out the feedback resistor
on U2 (see Figure \ref{fig:laserschematic}) with a wire, and then
turning on the laser pointer and moving the detector so that the
laser is centered on it.  With the feedback resistor on U2
shorted, the motor should be stationary with no voltage driving
it.  (The Gain of U2 is proportional to the value of the feedback
resistance.  If that resistance is zero, the gain of U2 is also
zero, meaning zero output voltage.)

When the detector is approximately in the right place, we remove
the wire that shorts out the feedback detector on U2.  Assuming
that your detectors are working properly, and that none of your
op-amps has burned out, one of four things happens.

\begin{enumerate}
\item{Nothing.  If the feedback gain is not large enough, the error
signal will not be large enough to make the motor move.  So,
assuming that everything else is working correctly, try increasing
the feedback gain, i.e., increase the value of the feedback
resistor on U2. }

\item{The laser swings off in the wrong direction.  Be careful about
this.  Stray laser beams can pose a safety problem.  It means that
your error signal has the wrong polarity. The easiest way to fix
this problem is to change the inputs on U1. So instead of
connecting the righthand detector to the inverting input and the
lefthand connector to the non-inverting input, switch them.  You
can also achieve the same thing by turning your detector over so
that left is now right, and vice versa. }

\item{The laser-pointing angle oscillates back and forth.
The feedback gain is too high.  Try reducing the gain
by using a smaller feedback resistor on U2.  Actually,
it is interesting to see this behavior.  You can induce
it by replacing U2's feedback resistor with a 0.1 $\mu$F
capacitor.  Try it.
}

\item{It works: the laser angle quickly changes to cause the
laser to point to the center of the detector.  When you get to
this point, it is fun to find out how good your locking system
really is.  Move the target and watch the laser follow it.  Move
it fast and slow and watch what happens.  Find out how close the
laser has to be to the center of the target during the initial
setup in order to lock the laser to the target.  Change the gain
(larger and smaller feedback resistors on U2) to see how it
affects the stability and speed of response of the lock. }

\end{enumerate}

\section{A Stabilizing Circuit for a \\ Michelson Interferometer}

With the rudiments of feedback control in hand, your students may
wish to try a more advanced application.  A few advanced projects,
such as frequency stabilizing a laser diode \cite{Libbrecht96} or
a He-Ne laser \cite{Jones93} and intensity stabilizing an LED
\cite{EDN}, are described in the literature.  In this section we
describe stabilizing a Michelson interferometer.

\begin{figure}
\includegraphics[angle=270,width=3.4in]{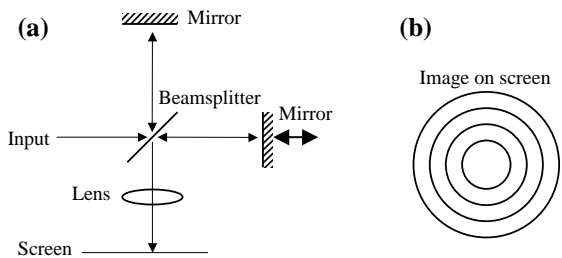} %figure 5
\caption{Optical layout for a Michelson interferometer.}
\label{fig:michelson}
\end{figure}

Probably the Michelson interferometer is familiar to all readers
\cite{Hecht}.  It is shown schematically in Figure
\ref{fig:michelson}a.  Light enters the interferometer at a beam
splitter.  Part of the wave is transmitted through the
beamsplitter, and part is reflected.  Each of these beams is
retroreflected by a mirror back to the beam splitter.  During the
round-trip from the beamsplitter to the mirror and back to the
beamsplitter, each beam accrues a phase relative to the input
beam.  If the round-trip distance is different for the two beams,
they also accrue a phase relative to each other.  When the two
beams are recombined on the beamsplitter, this relative phase can
produce constructive or destructive interference, making the
output bright or dark.

In the standard arrangement, the output of the interferometer is
projected onto a screen by a lens.  The output is a concentric set
of rings, alternately bright and dark, as shown in Figure
\ref{fig:michelson}b.  When one of the retroreflecting mirrors
moves either toward or away from the beamsplitter, the rings will
either collapse into or expand outward from the center of the
pattern.  If your mirrors vibrate, or if there are air currents in
the room, the ring position will fluctuate on the screen.  It is
this ring position on the screen that we want to control.

Controlling the ring position is quite similar to controlling the
laser-pointing angle.  It requires a small two-segment
photo-detector, with an active area roughly equal to the width of
one of the bright rings.  The analysis of the locking circuit is
identical to the laser pointer.  The detector has a ``left'' and a
``right.''  The difference of the signals from the left and right
sides of the detector indicate the position of the bright and dark
fringe relative to the center of the detector. This difference is
amplified and fed to a piezoelectric (PZT) crystal, which
translates one of the retroreflecting mirrors in the right
direction to move the edge of one ring to the center of the
detector.

\begin{figure}
\includegraphics[angle=270,width=3.4in]{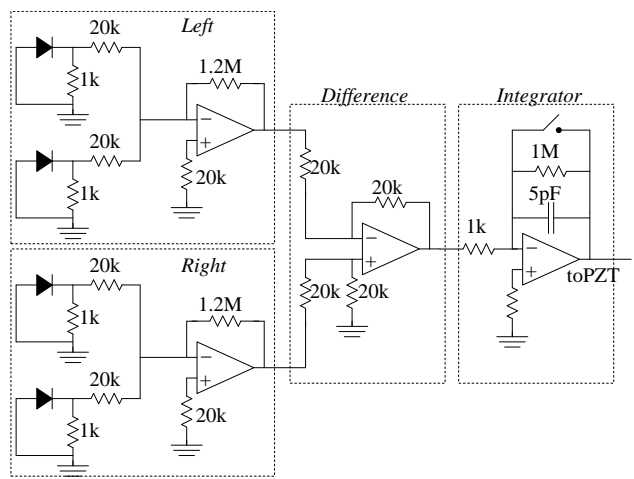} %figure 6
\caption{Schematic diagram of the electronics used to stabilize a
Michelson interferometer.  The op-amps are all the familiar
741's.} \label{fig:michelsonschematic}
\end{figure}

Our detector for this lab was a quadrant photodetector (Hamamatsu
S1557).  We mounted the detector onto a PC board and combined the
four quadrants into two pairs, so that the detector output had
only a ``left'' and ``right'' signal.  The electronics are shown
schematically in Figure \ref{fig:michelsonschematic}. The laser
intensity on the photodetector is much smaller in this lab
exercise than in the laser pointer experiment.  Accordingly, the
summing amplifiers provide some gain (50x).  The values you use in
your setting will depend on the intensity of the laser, the
efficiency of your detector, and so forth.

The difference signal is generated by taking the difference
between the left and right segments of the detector with a
difference amplifier, as shown in Figure
\ref{fig:michelsonschematic}.  The feedback amplifier follows the
difference amplifier, with either integral gain (shown in the
diagram) or proportional gain.

Not shown in the diagram is an optional offset circuit and
high-voltage amplifier.  Our PZT crystal (Thorlabs AE0203D04)
expands at roughly 0.08 $\mu$m/V, and has a maximum voltage rating
of 150 V.  It is intended to operate with a positive bias voltage.
Reversing the voltage on the device shortens its lifetime, but the
device can tolerate moderate negative voltages.  In our
experiment, we drive the PZT only to $\pm$10V.  In an optimized
circuit, the output would be shifted from $\pm$10V to range from
0V to 20V, and then amplified to range from 0V to 150V.  However,
these last steps make the laboratory exercise somewhat more
involved, and don't significantly increase the student's
understanding of closed-loop feedback control.

As before, the students place the detector in roughly the right
position with the feedback gain set to zero (i.e., the switch
across the capacitor in Figure \ref{fig:michelsonschematic}
closed), and then open the switch to engage the ``lock'' to the
fringe.  As with most labs, the students' first attempt at locking
probably won't work.  The list of things to check are roughly the
same as those for the laser pointer, and the reader is referred to
the list above.

The goal is to have the error signal driving the PZT range over
±10 Volts as the detector traverses one ring. If it does not, the
feedback gain in the difference amplifier may need to be
increased.  In Figure 6 the difference amplifier gain is unity.
Try increasing this gain by increasing the 20 k-ohm feedback
resistor.  Conversely, if the gain is too high, the output of the
integrator will saturate at the supply voltage. The 1M-ohm
resistor in parallel with the capacitor between the negative input
and the output of the integrator is used to keep the integrator
from saturating due to inherent offset currents in the 741s. If
the circuit refuses to lock, this may mean that the error signal
going to the PZT has the wrong polarity.  As with the laser
pointer, the easiest way to fix this problem is to reverse the
inputs to the difference amplifier or to turn the detector over,
so left is now right and vice versa.

When the lock works, students can move the detector and watch the
fringe pattern change.  As the detector moves toward the center of
the fringe pattern, the fringes will also collapse toward the
center. When the detector moves away from the center of the fringe
pattern, the fringe pattern will also expand away from the center.
Of course, the PZT can only expand roughly one $\mu$m with these
voltages, so this effect can be observed only over a relatively
limited range.  The switch can be momentarily closed to reset the
signal to the PZT to zero in order to continue locking to the same
edge as the detector is moved further.

\section{Acknowledgements}

EFM would like to thank M. Noel for his help with the Michelson
interferometer experiment.  This work is supported in part by a
grant from the Research Corporation and from the National Science
Foundation under Grants No. PHY-9985027 and PHY-9623569.


\begin{thebibliography}{}
\bibitem{lock}
When a system output is tightly regulated to a desired output, the
system is said to be ``locked'' to the output.
\end{thebibliography}
\end{document}